\documentclass[sigconf]{acmart}

\AtBeginDocument{%
  }

\setcopyright{acmlicensed}
\copyrightyear{2024}
\acmYear{2024}
\acmDOI{XXXXXXX.XXXXXXX}

\acmConference[Conference acronym 'XX]{Make sure to enter the correct
  conference title from your rights confirmation emai}{June 03--05,
  2018}{Woodstock, NY}
\acmISBN{978-1-4503-XXXX-X/18/06}

\usepackage{caption}
\usepackage{subcaption}

\usepackage{tabularx}
\usepackage{booktabs}

\usepackage[textsize=footnotesize]{todonotes} 
\usepackage[subtle]{savetrees} % Less legal, but admissable.
\newcommand{\shrink}{\vspace{-0.5\baselineskip}}
\newcommand{\sshrink}{\vspace{-0.25\baselineskip}}
\newcommand{\mypar}[1]{\medskip\pagebreak[3]\noindent\textbf{#1}~}
\renewcommand{\mypar}[1]{\smallskip\pagebreak[3]\noindent\textbf{#1}~}
\begin{document}

\title{Interpretability Analysis of Domain Adapted Dense Retrievers}

\author{G\"{o}ksenin Y\"{u}ksel}
\email{goksenin.yuksel@student.uva.nl}
\orcid{0009-0005-5045-0522}
\affiliation{%
  \institution{University of Amsterdam}
  \city{Amsterdam}
  \country{Netherlands}
}

\author{Jaap Kamps}
\email{kamps@uva.nl}
\orcid{0000-0002-6614-0087}
\affiliation{%
  \institution{University of Amsterdam}
  \city{Amsterdam}
  \country{Netherlands}
}

\renewcommand{\shortauthors}{Trovato et al.}

\begin{abstract}
Dense retrievers have demonstrated significant potential for neural information retrieval; however, they exhibit a lack of robustness to domain shifts, thereby limiting their efficacy in zero-shot settings across diverse domains. 
Previous research has investigated unsupervised domain adaptation techniques to adapt dense retrievers to target domains. However, these studies have not focused on explainability analysis to understand how such adaptations alter the model's behavior.
In this paper, we propose utilizing the integrated gradients framework to develop an interpretability method that provides both instance-based and ranking-based explanations for dense retrievers. 
To generate these explanations, we introduce a novel baseline that reveals both query and document attributions. 
This method is used to analyze the effects of domain adaptation on input attributions for query and document tokens across two datasets: the financial question answering dataset (FIQA) and the biomedical information retrieval dataset (TREC-COVID).
Our visualizations reveal that domain-adapted models focus more on in-domain terminology compared to non-adapted models, exemplified by terms such as "hedge," "gold," "corona," and "disease." 
This research addresses how unsupervised domain adaptation techniques influence the behavior of dense retrievers when adapted to new domains. 
Additionally, we demonstrate that integrated gradients are a viable choice for explaining and analyzing the internal mechanisms of these opaque neural models.
\end{abstract}

\if 0 % Skip this for submission



\fi % Skip for submission

\maketitle

\shrink
\section{Introduction}

%One promising advancement in information retrieval (IR)  literature is the development of neural dense retrieval methods. 
One of the main recent achievements in information retrieval (IR) is the development of neural dense retrieval methods. 
These methods are extremely fast and do not entail the memory and computational overhead associated with widely used other neural methods such as cross-encoder and late interaction models~\cite{dense-survey}. However, dense retrieval models face the challenging task of independently mapping inputs to a meaningful vector space, making them highly sensitive to domain shifts~\cite{thakur-etal-2021-augmented}. This non-robustness to domain shifts impedes their application in zero-shot settings~\cite{beir, zero-shot}, posing challenges in real-world applications where access to extensive and domain-specific training data is limited~\cite{ma-etal-2021-zero}. Methods known as “domain adaptation” have been developed to address this issue.

To tackle the domain shift problem with dense retrievers, previous work has fine-tuned these models on target datasets using unsupervised and supervised learning objectives~\cite{QGen, wang2021gpl, thakur-etal-2021-augmented}. Supervised methods utilize labeled data to further fine-tune these models in novel domains, which is not possible for every task or domain of IR research due to the costs and difficulty of obtaining human-annotated labels~\cite{xin-etal-2022-zero}. In contrast, unsupervised methods assume only the availability of the target corpus~\cite{xin-etal-2022-zero, QGen, wang2021gpl}, employing pre-training objectives~\cite{gururangan-etal-2020-dont, lee-etal-2019-latent, liu-etal-2021-fast} or pseudo-labeling~\cite{wang2021gpl, QGen} to fine-tune the pre-trained models in a new domain without requiring labeled data. Notably,~\citet{wang2021gpl} found that pre-training objectives alone do not enhance the out-of-domain performance of the adapted model.

% Moved up as teaser on P1.
\begin{table}[!t]
    \caption{Retrieval effectiveness of domain adaptation.}
    \label{tab:eff}
    \vspace{-2ex}
    %\centering
    \setlength\tabcolsep{0pt} % let LaTeX figure out intecol. whitespace
    \begin{tabular*}{\columnwidth}{@{\extracolsep{\fill}} lcccc}
    \toprule
    \bf Collection & \multicolumn{4}{c}{\bf NDCG@10} \\
    \cmidrule{2-5}
    & \bf Baseline & \bf GLP & \bf Absolute & \bf Percentage \\
    \midrule
    TREC-COVID 
    & 0.6510 & 0.7160 & $+$0.0650 & $+$9.98\% 
    \\
    FIQA 
    & 0.2670 & 0.3680 & $+$0.1010 & $+$37.83\% 
    \\
    \bottomrule
    \end{tabular*}
\end{table}

%Despite the importance and interest in domain adaptation, previous work
Prior work on domain adaption focuses on retrieval effectiveness 
% Add teaser table here.
(e.g. Table~\ref{tab:eff} discussed below), 
but has not addressed the interpretability of domain-adapted models. Moreover, no model-introspective analysis was conducted to understand changes in the models' inner workings before and after domain adaptation. Given the inherent opacity and lack of transparency of these neural models, interpretability analysis is crucial to understand the effect of domain adaptation on such dense retrievers.
%
%To address this issue, we 
This paper conducts initial interpretability analysis experiments, and proposes to use of the Integrated Gradients (IG)~\cite{integrated-gradients} framework to develop an interpretability method for dense retrievers, providing both instance-based and ranking-based explanations. Subsequently, we apply this method to domain-adapted and non-domain-adapted dense retrievers to assess the differences in their behavior using input-based attributions.
We set out to answer two research questions. \textsl{(i) How can Integrated Gradients be utilized in the dense retriever setting? (ii) How does domain adaptation influence the input attributions of the models?}

In this paper, we first demonstrate that IG is a viable approach for interpreting dense retrievers. We then utilize the proposed interpretability method to analyze input attribution differences in the FIQA and TREC-COVID datasets, which represent two distinct domains in IR: financial question answering and biomedical information retrieval. Our visualizations reveal that the domain-adapted model concentrates more on in-domain terminology, and title which the unadapted model tends to overlook.

\shrink 
\section{Related Research}

\sshrink 
\subsection{Unsupervised Domain Adaptation} % in Dense Retrievers}

\textbf{Query Generation (QG):} QG methods construct synthetic training data by using documents from the target domain to generate corresponding (pseudo) queries, aiming to augment the training data with queries that fit the target domain. QGen~\cite{QGen} trains an auto-encoder in the source domain to generate synthetic questions from a target domain document. They use binary-level relevancy labels to train the networks on generated query-document pairs. Similarly, GPL~\cite{wang2021gpl} generates synthetic queries with a pre-trained T5 model but uses cross-encoders to label the relevancy of generated query-document pairs. This method extends and over performs the QGen method by replacing binary relevance with continuous labels ranging from -inf to inf.

\textbf{Knowledge Distillation (KD):} KD is a commonly used strategy in the dense retriever setting, which utilizes a powerful model as the teacher model to improve the capabilities of the student models~\cite{KD-DR}. It has been found that such a technique can improve out-of-domain performance as well. GPL~\cite{wang2021gpl} and AugSBERT~\cite{thakur-etal-2021-augmented} use cross-encoders to annotate unlabeled synthetic query-doc pairs. Later, this knowledge is distilled into the dense retriever by training the model on generated labels. Different from the above methods, SPAR~\cite{chen-etal-2022-salient} proposes to distill knowledge from BM25 to the dense retriever model to integrate sparse retrieval.

\sshrink  
\subsection{Interpretability methods for IR}

Interpretability of ranking models focuses on building models that can either be analyzed for interpretability in a post-hoc fashion or are interpretable by design.

Post-hoc interpretability methods explain the decisions of already trained machine learning models. These approaches are either model-agnostic, where the interpretability approach has no access to the trained model parameters~\cite{shap, lime}, or model-introspective, with full access to the parameters of the underlying model~\cite{white1, integrated-gradients}. There exists a lot of work on model-introspective analysis of neural models for different tasks. These analyses use different methods like probing tasks, attention weights, or state activations~\cite{explainable-ir}. A recent dominant class of model-introspective explanation outputs feature attributions. Most of these approaches utilize gradient-based attribution methods~\cite{explainable-ir}.

To our knowledge, only \citet{people-used-bert} used IG to obtain feature attributions for a BERT-based cross-encoder model. In their experiments, they used an empty query baseline and an empty document baseline, which are padding tokens. 

\shrink 
\section{Methodology}
\sshrink

\mypar{Dense Retriever}
In this paper, we use pre-indexing of the documents. The pre-indexed document embeddings are used to retrieve top-K passages using dot product similarity between document and query embeddings. The query and document embeddings are obtained using a DistilBERT model, which has already been fine-tuned on MSMARCO to retrieve relevant documents to a query~\cite{msmarco}.

We use the dense retriever models that are tuned to work with the dot product as their similarity measure. We utilize open sourced GPL models and use SentenceTransformer~\cite{reimers-2019-sentence-bert} framework. For all the GPL models, maximum sequence length is set to 350. We use mean pooling over output token embeddings, disregarding the special tokens such as [CLS] and [SEP]. For TREC-COVID, and FIQA dataset we evaluate provided the models, and asses the performance improvement on the corresponding test sets.

\mypar{Baseline}
As stated by \citet{integrated-gradients}, a good baseline should give a score of zero and should convey an empty signal. Inspired by the recent work in cross-encoder explanations, we use the [PAD] token to create our baseline. However, we deviate from them by introducing a new method to calculate the query and document attributions.
To calculate the query token attributions, we replace the query tokens with the [PAD] tokens and leave the document tokens untouched. Then, we replace the document tokens with [PAD] tokens and leave the query tokens untouched. We run the Integrated Gradient analysis for both the query and the document baselines. This method calculates the input attributions for both the query and the document tokens.

\mypar{Ranking Analysis}
The aforementioned method only works for instance-based explanations. However, in the information retrieval setting, we are also interested in the explainability of the document ranking. For this task, we select the top 25 documents retrieved by the model for a specific query.
We aggregate the token attributions over the top retrieved documents by summing them up to generate the overall attributions of a token. This way, we get the most important tokens for the ranking. The tokens that appear often in the top rankings and contribute positively get a higher attribution score for the ranking overall. We use word cloud visualization, the word sizes are determined by the summed attribution over 25 documents.

\begin{figure}[t]
  \centering
  \begin{subfigure}[t]{\linewidth}
  \includegraphics[width=\linewidth]{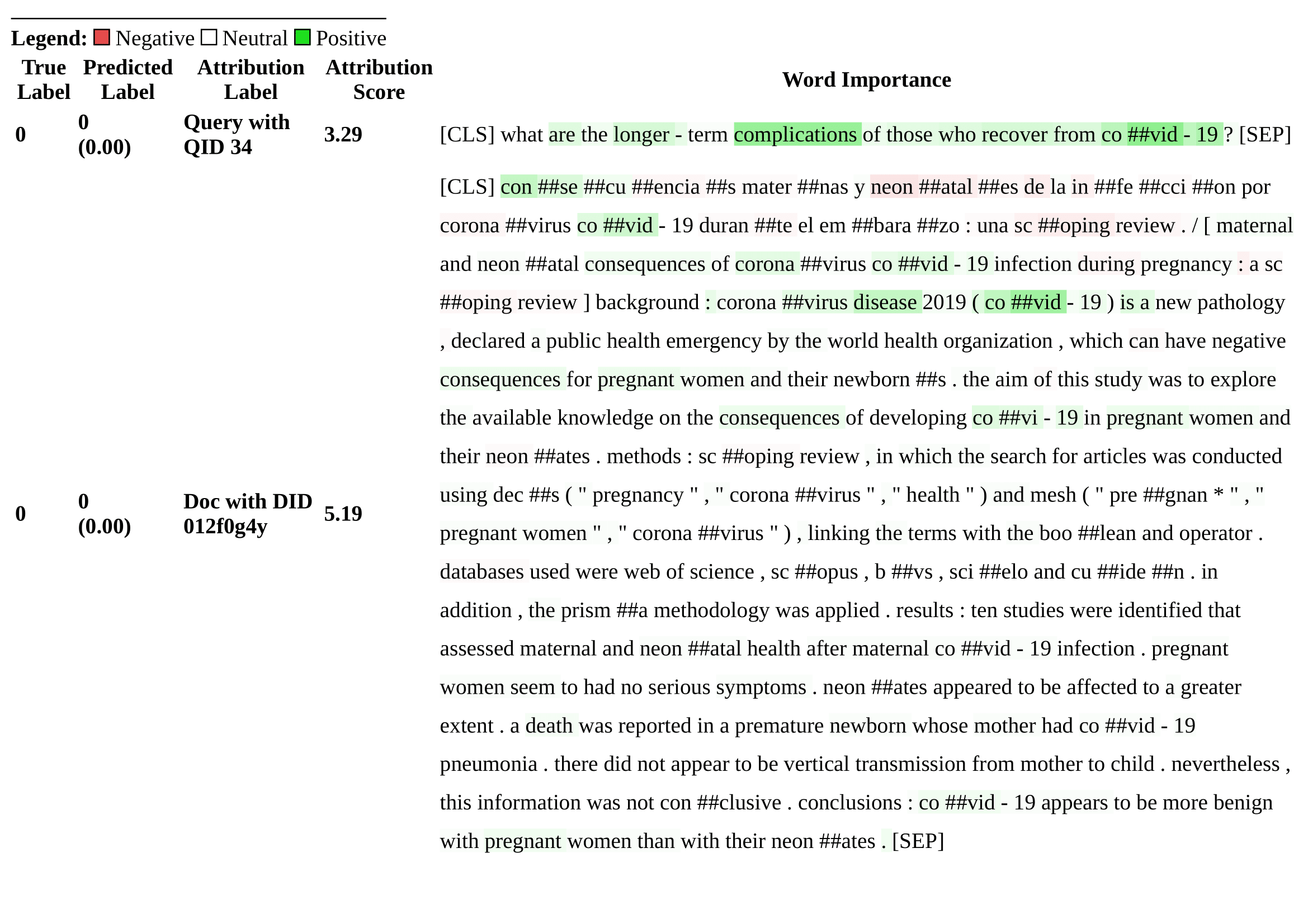}
     \caption{Instance Based}
     \label{fig:instance-based-trec}
      \Description{Instance Based HTML Representation of the input attributions extracted from Integrated Gradients method for TREC-COVID query and document.}
      \end{subfigure}
   \hfill
  \begin{subfigure}[t]{0.48\linewidth}
      \includegraphics[width=\linewidth]{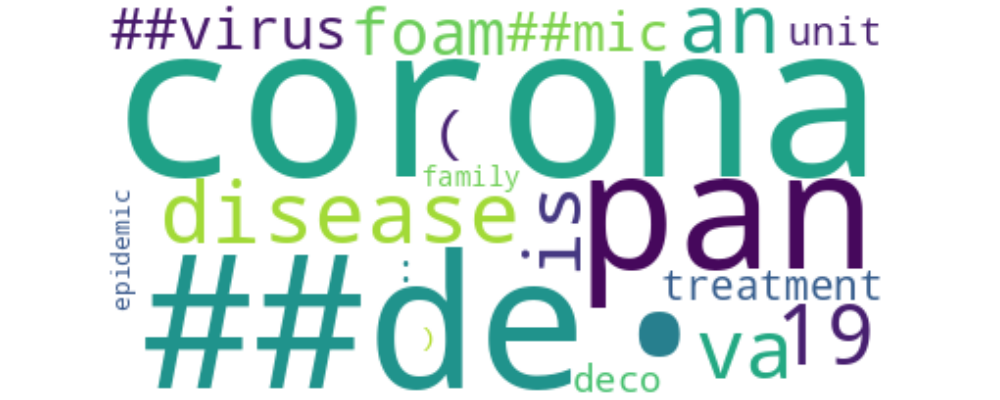}
      \caption{Ranking Based Positive Attribution}
      \label{trec-base-positive-ranking}
      \Description{Ranking Based Positive wordcloud Representation of the input attributions extracted from Integrated Gradients method for TREC-COVID query and top ranked documents.}
  \end{subfigure}
  \hfill
  \begin{subfigure}[t]{0.48\linewidth}
        \includegraphics[width=\linewidth]{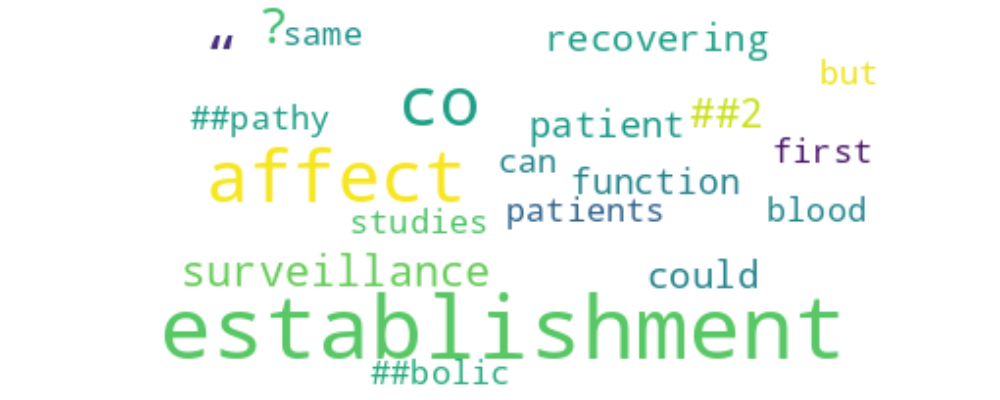}
        \caption{Ranking Based Negative Attribution}
        \label{trec-base-negative-ranking}
      \Description{Ranking Based Negative wordcloud Representation of the input attributions extracted from Integrated Gradients method for TREC-COVID query and top ranked documents.}
  \end{subfigure}
  \caption{Attribution analysis for random query and relevant document for TREC-COVID. The model used was GPL/msmarco-distilbert-margin-mse. The word sizes are determined by the summed attribution over top ranked 25 documents}
  \Description{A woman and a girl in white dresses sit in an open car.}
  \label{fig:first-trec}
\end{figure}

\begin{figure}[t]
  \centering
  \begin{subfigure}[t]{\linewidth}
    \includegraphics[width=\linewidth]{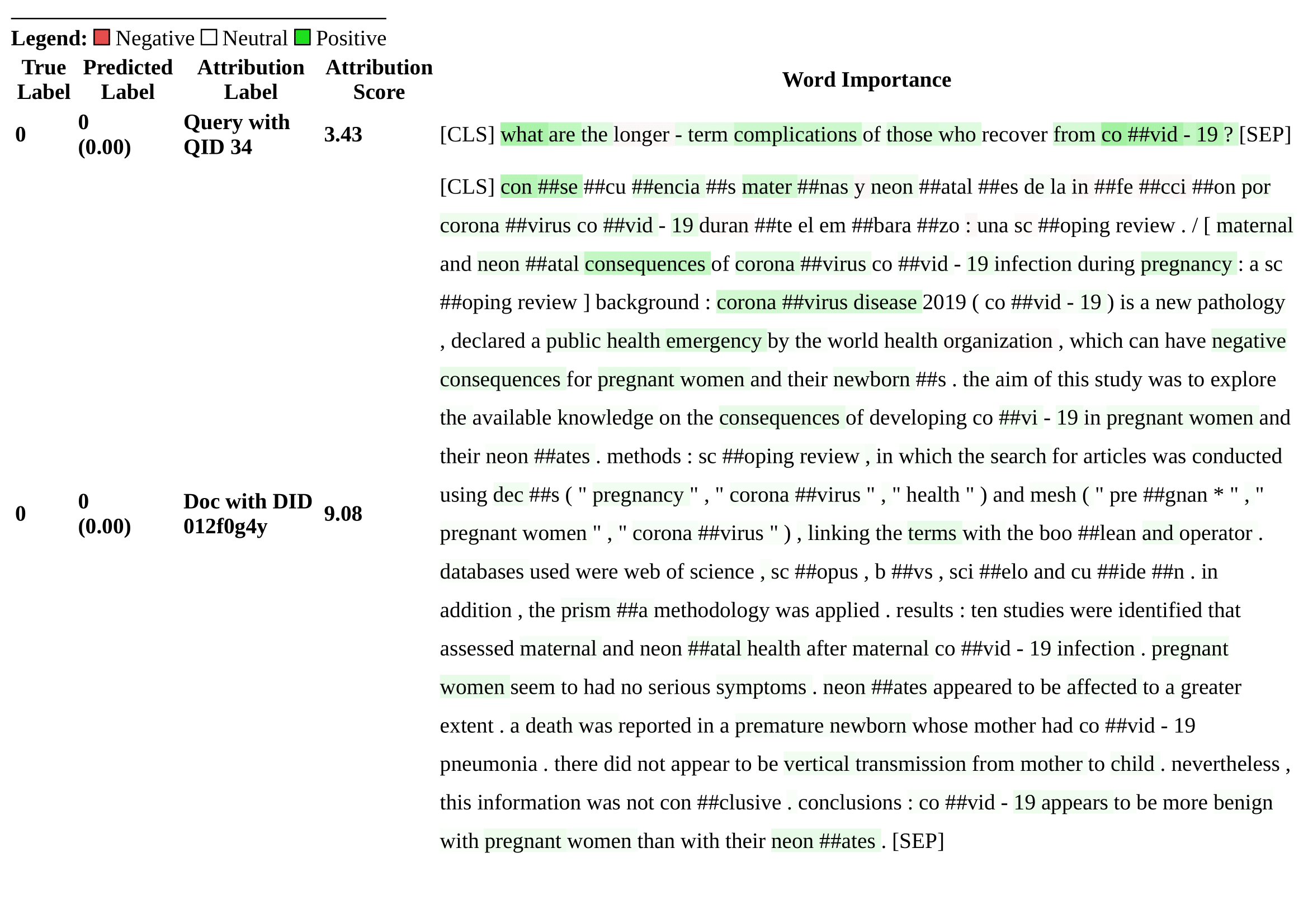}
    \caption{Instance Based}
    \label{trec-instance-adapted}
      \Description{Instance Based HTML Representation of the input attributions extracted from Integrated Gradients method for TREC-COVID query and document for domain adapted model.}  \end{subfigure}
  \hfill
  \begin{subfigure}[t]{0.48\linewidth}
      \includegraphics[width=\linewidth]{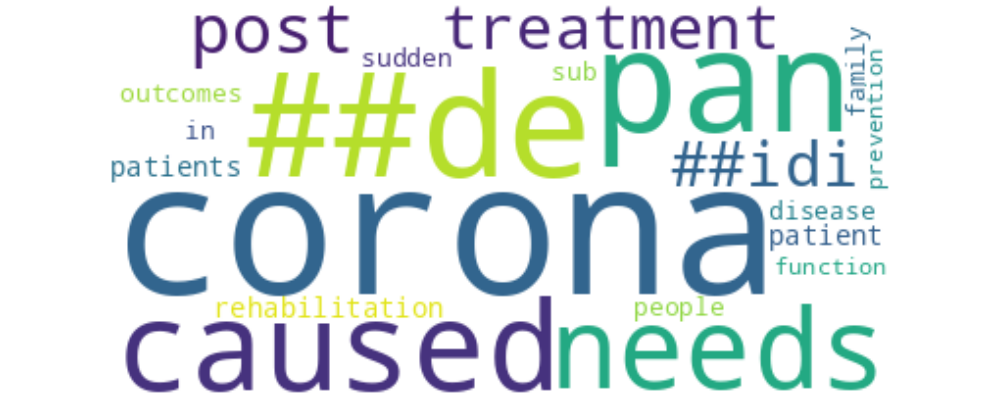}
      \caption{Ranking Based Positive Attribution}
      \label{trec-adapted-positive-ranking}
      \Description{Ranking Based Positive wordcloud Representation of the input attributions extracted from Integrated Gradients method for TREC-COVID query and top ranked documents for domain adapted model.}
  \end{subfigure}
  \hfill
  \begin{subfigure}[t]{0.48\linewidth}
    \includegraphics[width=\linewidth]{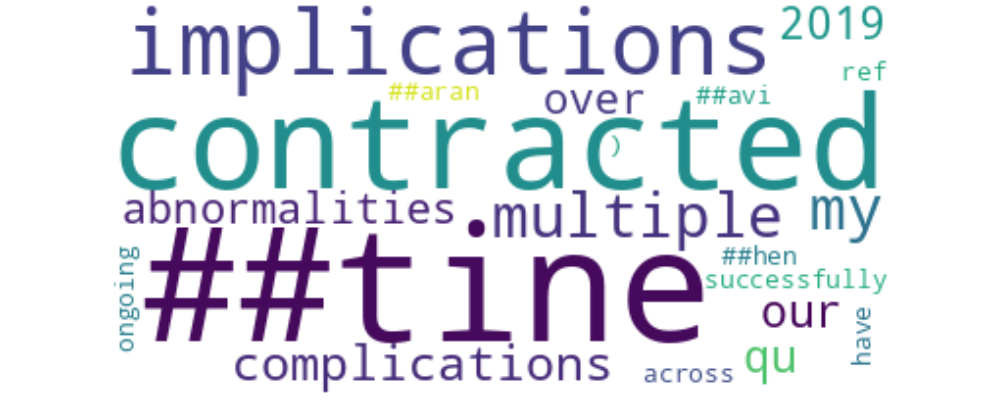}
    \caption{Ranking Negative Attribution}
    \label{trec-adapted-negative-ranking}
      \Description{Ranking Based Negative wordcloud Representation of the input attributions extracted from Integrated Gradients method for TREC-COVID query and top ranked documents for domain adapted model.}
  \end{subfigure}
  \caption{Attribution analysis for random query and relevant document for TREC-COVID. The model used was GPL/trec-covid-msmarco-distilbert-gpl. The word sizes are determined by the summed attribution over top ranked 25 documents}
  \label{fig:trec-adapted}
\end{figure}

\mypar{Title Attribution Analysis}
For TREC-COVID data, title is appended in the beginning of the document, and then the domain adaptation is performed. FIQA does not have a title attribute. Also for MSMARCO models, this is not the case as they do not include a title.

Hence for TREC-COVID, we can test for the difference in total attribution for the title to understand the domain adaptation affect. 
For each query, we randomly select a relevant document, and compute the document token attributions for both base and domain adapted model. Later, we sum the attribution for title tokens.

\shrink 
\section{Results}

Our experiments, 
shown in Table~\ref{tab:eff} before, 
confirm the effectiveness of domain adaption observed in earlier research, on TREC-COVID NDCG@10 improves from 65.1 to 71.6 (+6.5 abs., +10\%) and on FIQA from 26.7 to 36.8 (+10.1 abs., +38\%). With the proposed method, we now analyze if we can interpret how the domain adaptation is affecting the model.

\shrink
\subsection{Input Attribution}
\begin{figure}[!t]
  \centering
  \begin{subfigure}[t]{\linewidth}    
      \includegraphics[width=\linewidth]{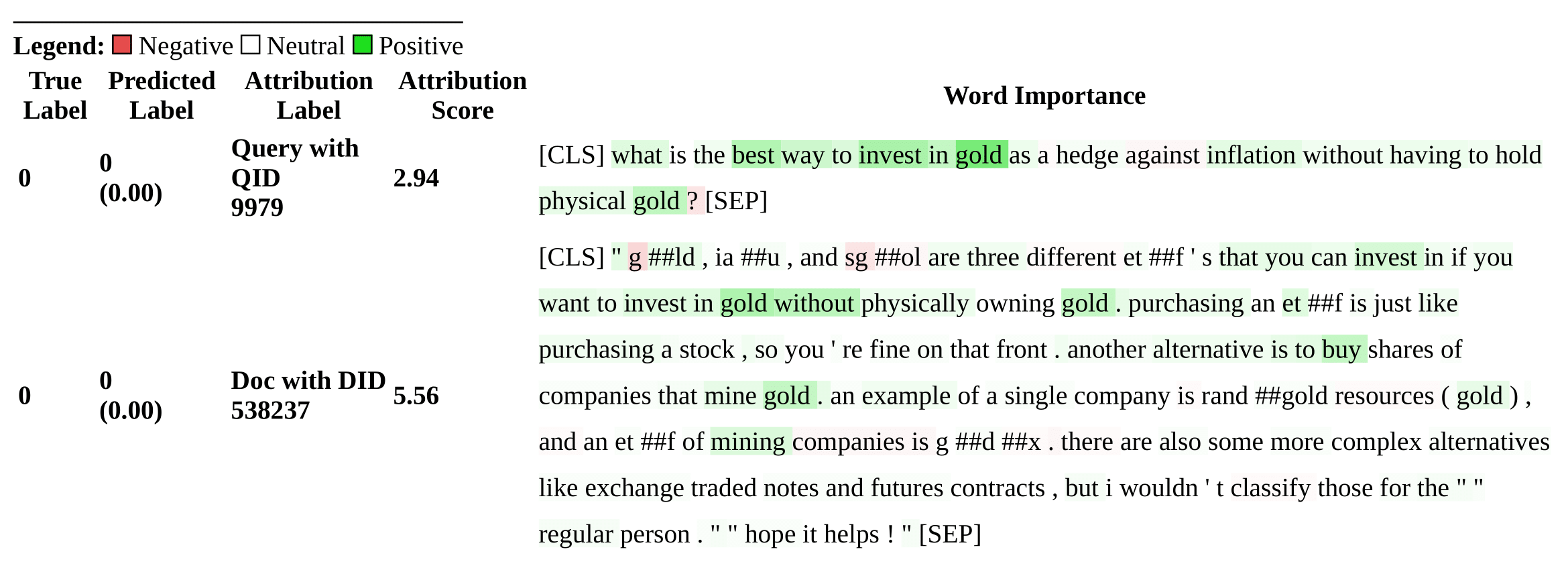}
      \caption{Instance Based}
      \label{fig:instance-based-fiqa}
      \Description{Instance Based HTML Representation of the input attributions extracted from Integrated Gradients method for FIQA query and document.}
  \end{subfigure}
  \hfill
  \begin{subfigure}[t]{0.48\linewidth}
      \includegraphics[width=\linewidth]{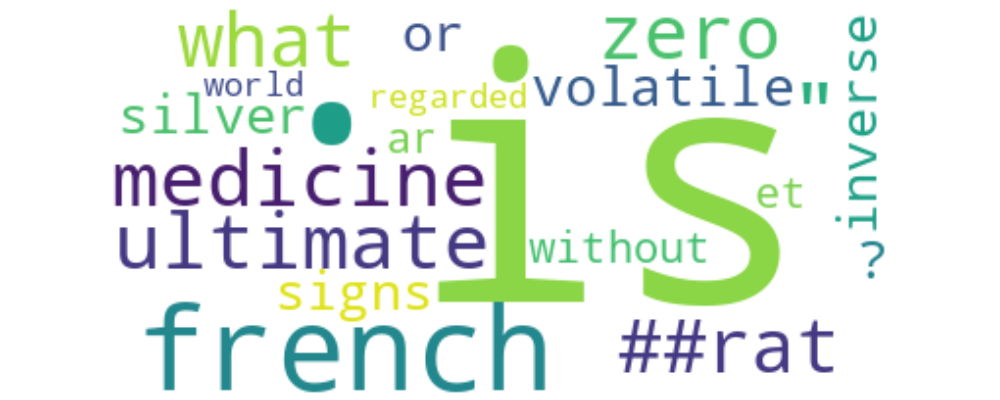}
      \caption{Ranking Based Positive Attribution}
      \label{fiqa-based-positive-ranking}
      \Description{Ranking Based Positive wordcloud Representation of the input attributions extracted from Integrated Gradients method for FIQA query and top ranked documents. The word sizes are determined by the summed attribution over top ranked 25 documents}
      \end{subfigure}
  \hfill   
  \begin{subfigure}[t]{0.48\linewidth}
      \includegraphics[width=\linewidth]{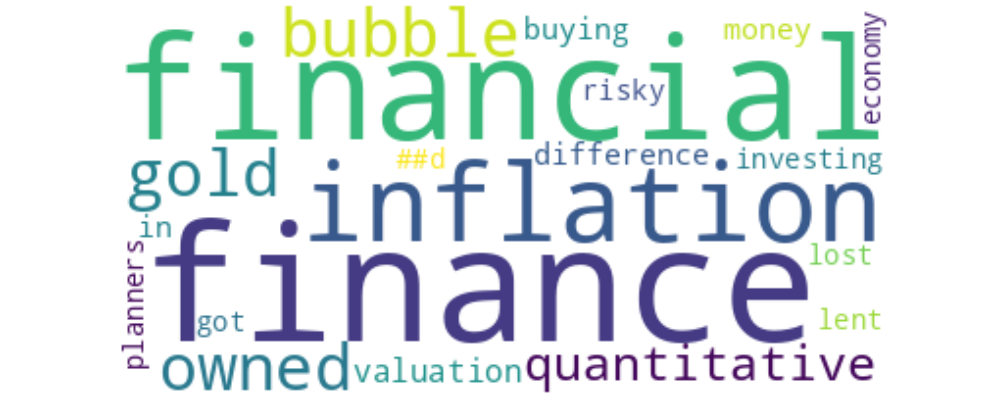}
      \caption{Ranking Based Negative Attribution}
      \label{fiqa-based-negative-ranking}
      \Description{Ranking Based Negative wordcloud Representation of the input attributions extracted from Integrated Gradients method for FIQA query and top ranked documents.}
    \end{subfigure}
    \caption{Attribution analysis for random query and relevant document for FIQA. The model used was GPL/msmarco-distilbert-margin-mse. The word sizes are determined by the summed attribution over top ranked 25 documents}
    \label{fig:first-fiqa}
\end{figure}

Figures~\ref{fig:first-trec} and~\ref{fig:first-fiqa} show the attribution analysis for the baseline model.

Figures \ref{fig:instance-based-trec} and \ref{fig:instance-based-fiqa} display the positive and negative attributions for both query and document pairs in models trained on MSMARCO using DistilBERT. Positive attributions enhance the similarity between the query and the document, while negative attributions lowers it.

In Figure \ref{fig:instance-based-trec}, the query tokens ["complications," "co," "\#\#vid"] contribute positively. Additionally, the document tokens ["co," "\#\#vid," "disease"] also contribute positively, whereas ["neon"] contributes negatively.

In Figure \ref{fig:instance-based-fiqa}, the query tokens ["invest," "gold," "best"] contribute positively, while ["against," "?"] contribute negatively. Furthermore, the document tokens ["gold," "without," "buy"] contribute positively, whereas ["g," "sg"] contribute negatively.

For both queries, the DistilBERT model effectively matches query terms with document terms. Tokens that appear in both the query and the document receive high positive attribution scores. As expected, we observe no attributions for the [CLS] and [SEP] tokens.

\mypar{Ranking Based Attribution} 
Figures \ref{trec-base-positive-ranking}, \ref{trec-base-negative-ranking}, \ref{fiqa-based-positive-ranking}, and \ref{fiqa-based-negative-ranking} depict the word clouds of positive and negative attributions revealed by the proposed method in a ranking scenario.

Despite the query not explicitly mentioning ["corona," "disease"], Figure \ref{trec-base-positive-ranking} illustrates that these are important words identified by the model. Another notable finding is the appearance of "complications" in the negative attribution word cloud shown in Figure \ref{trec-base-negative-ranking}, which is a term present in the query text.

For the FIQA dataset, Figures \ref{fiqa-based-positive-ranking} and \ref{fiqa-based-negative-ranking} show that the model assigns positive attributions to document tokens ["what," "is," "french"] and negative attributions to ["financial," "finance"]. Additionally, "gold" appears among the negatively contributing terms. 

\subsection{Domain Adaptation}
\begin{figure}[t]
  \centering
  \begin{subfigure}[t]{\linewidth}
      \includegraphics[width=\linewidth]{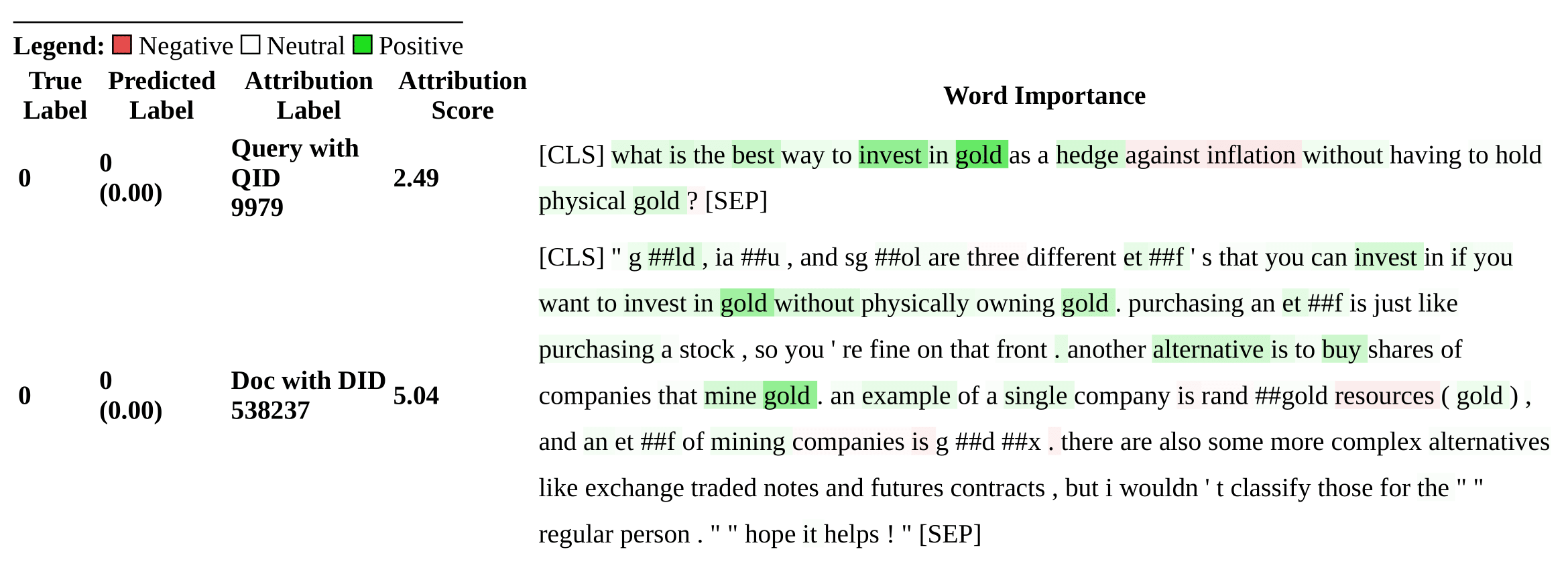}
      \caption{Instance Based}
      \label{fig:instance-adapted-fiqa}
      \Description{Instance Based HTML Representation of the input attributions extracted from Integrated Gradients method for FIQA query and document for domain adapted model.}
  \end{subfigure}
  \hfill
  \begin{subfigure}[t]{0.48\linewidth}
      \includegraphics[width=\linewidth]{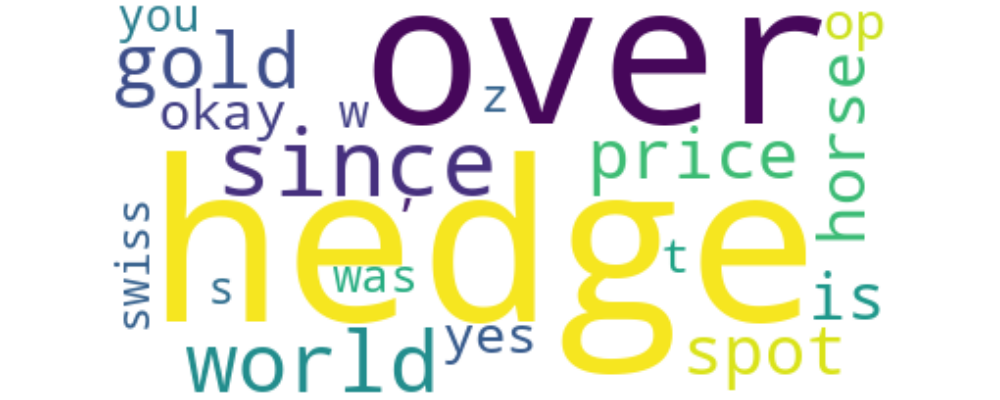}
      \caption{Ranking Based Positive Attribution}
      \label{fiqa-adapted-positive-ranking}
      \Description{Ranking Based Positive wordcloud Representation of the input attributions extracted from Integrated Gradients method for FIQA query and top ranked documents for domain adapted model.}
      \end{subfigure}
    \hfill
    \begin{subfigure}[t]{0.48\linewidth}
      \includegraphics[width=\linewidth]{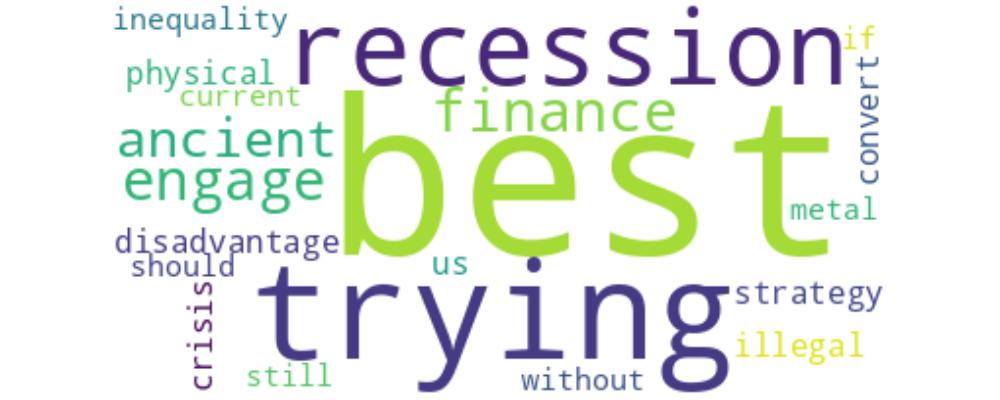}
      \caption{Ranking Based Negative Attribution}
      \label{fiqa-adapted-negative-ranking}
      \Description{Ranking Based Negative wordcloud Representation of the input attributions extracted from Integrated Gradients method for FIQA query and top ranked documents for domain adapted model.}
      \end{subfigure}
  \caption{Attribution analysis for random query and relevant document for FIQA. The model used was GPL/fiqa-msmarco-distilbert-gpl. The word sizes are determined by the summed attribution over top ranked 25 documents}
  \label{fig:fiqa-adapted}
\end{figure}

Figures~\ref{fig:trec-adapted} and~\ref{fig:fiqa-adapted} show the attribution analysis after domain adaptation. 
%Figures~\ref{trec-instance-adapted} and~\ref{fig:instance-adapted-fiqa} display the attribution analysis after domain adaptation. 

In Figure \ref{trec-instance-adapted}, the query tokens ["what," "co," "\#\#vid"] contribute positively. Additionally, the document tokens ["pregnancy," "consequences," "corona"] contribute positively. The query contribution of "complications" has decreased compared to the non-domain-adapted model, while the contribution of "what" has increased. Furthermore, the domain-adapted model assigns more importance to the first sentence, which is the title of the paper. The attribution for document tokens ["corona," "consequences," "pregnant"] has also increased.

In Figure \ref{fig:instance-adapted-fiqa}, the query tokens ["invest," "gold," "hedge"] contribute positively, whereas ["against," "inflation"] contribute negatively. Additionally, the document tokens ["gold," "invest," "buy"] contribute positively, while ["resources," "is"] contribute negatively. The query contribution of "inflation" has decreased after domain adaptation, whereas "best" has increased. Moreover, the attribution for document tokens ["gold," "alternative," "g"] has increased, whereas ["resources," "without"] has decreased.

Both models display positive attributions for the input tokens in the sentence, "you can invest in if you want to invest in gold without physically owning gold."

\mypar{Ranking}
For TREC-COVID, Figures \ref{trec-base-positive-ranking} and \ref{trec-adapted-positive-ranking} illustrate that both the non-domain-adapted model and the domain-adapted model place emphasis on "corona" and "pan." A notable finding is the increase in attribution to "treatment" and the appearance of "post" in the domain-adapted model.

Regarding negative attributions, Figures \ref{trec-base-negative-ranking} and \ref{trec-adapted-negative-ranking} show that the non-domain-adapted model focuses more on "establishment," "affect," and other non-related query words. In contrast, the domain-adapted model focuses on "contracted" and "implications." In both cases, they do not assign negative attributions to query-related terms. Additionally, the domain-adapted model identifies "2019" as a negatively contributing token to the ranking.

For FIQA, Figures \ref{fiqa-based-positive-ranking} and \ref{fiqa-adapted-positive-ranking} reveal that the domain-adapted model places greater emphasis on "hedge," "gold," and "over." Conversely, the non-domain-adapted model places more emphasis on "is" and "french." Furthermore, Figures \ref{fiqa-based-negative-ranking} and \ref{fiqa-adapted-negative-ranking} show that the non-domain-adapted model assigns high negative attribution to "finance" and "financial," whereas the domain-adapted model assigns high negative attribution to "best" and "trying." In this scenario, the domain-adapted model focuses more on query-related terms such as "hedge" and "gold" compared to the non-domain-adapted model.

\begin{figure}[!t]
    \centering
    \includegraphics[width=0.65\linewidth]{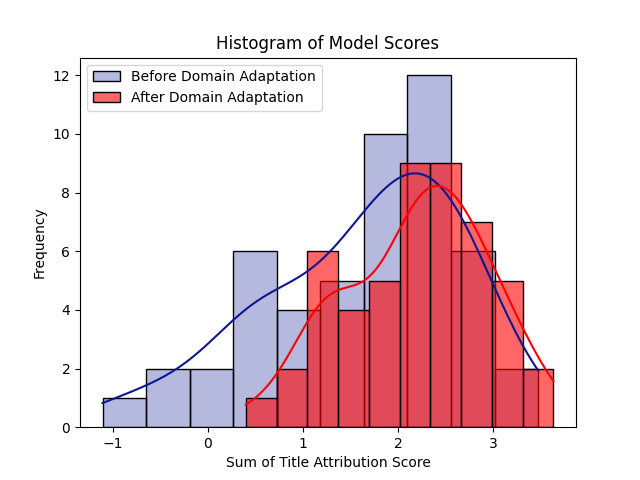}
    \caption{Sum of title attribution scores for TREC-COVID}
    \label{fig:title}
\end{figure}

\mypar{Title attributions}
Figure \ref{fig:title} depicts that domain adapted model puts positive attribution to the title compared to base model. The base model puts significantly negative attribute to the title. This quantitative finding is consistent with the qualitative observations above.

\shrink 
\section{Conclusions and Discussion}
\sshrink

This paper proposes to use Integrated Gradients for instance-based and ranking-based explanations of dense retrievers. We find that Integrated Gradients is a feasible choice for the interpretability of dense retriever models. Our initial findings show that dense retriever models are capable of both soft matching and term matching as expected. For example, even though the query may not contain the word "volatile," documents containing "volatile" are ranked higher, with a positive attribution score assigned to the term. Furthermore, we observe that negative attributions are also applicable in the dense retriever setting, with certain tokens contributing negatively to the similarity score. As illustrated in Figure~\ref{fig:first-trec}, these are typically non-relevant tokens for the query.

 We find that domain-adapted models tend to place more positive attributions on domain-specific vocabulary, such as "corona" and "hedge," compared to the baseline model. Additionally, the domain-adapted model better captures the dependencies of document terms like "treatment" and "rehabilitation," which are relevant to the query but not explicitly stated in the query terms, compared to the non-domain-adapted model. Moreover, the positive attribution given to the title by the domain-adapted model in the TREC-COVID domain suggests that the model relies more on the document title to generate similarity scores compared to the non-domain-adapted model. This behavior may be influenced by the domain adaptation process in GPL models, where the title is concatenated at the beginning of the document~\cite{wang2021gpl}. Thus, the model may learn to focus more on the beginning of the document, recognizing the title as an important element for retrieving research papers related to the query.
\shrink % Still too much space...
\appendix%
\section{Limitations}

Interpretable and explainable neural ranking models is an important, but also very hard to study problem in IR. Our initial experiments in this paper merely aim to raise interest in, and show the viability of, interpretability analysis of dense retrievers.
%
%However, there are multiple domain adaptation methods, each with different approaches to adapting the model. Furthermore, the conclusions for the second research question are based on %anecdotal evidence from two random queries. 
%This is due to lack of quantitative ways of %empirical analysis method for 
As there is no standard quantitative way of evaluating attributions produced by IG, we opted for a deep qualitative analysis of a small sample of queries.  We plan to considerably expand this in future work. 
%We also considered one domain adaptation method. 
We also plan to compare the IG method to other interpretability methods, %to reach better conclusions, 
such as~\cite{shap} or~\cite{lime}, in future research.

To gain a comprehensive understanding of the white/black box model, we also want to expand the analysis to the global attributions of models. However, IG is not capable of producing such global explanations.   %Additionally, we did not discuss the full ranking but focused on attributions in the top 25 ranked documents. Therefore, our methodology does not address lower-ranked documents and their explanations. Understanding the attributions for lower-ranked documents is crucial for more comprehensible and reliable IR research. 
We aim to extend this our research beyond the top ranked documents, 
as understanding the attributions for lower-ranked documents is crucial for more comprehensible and reliable IR research.
Moreover, it is possible to analyze the most influential document tokens for a query similarity by computing document token attributions using all the documents in the corpus.  We plan to conduct such very computationally demanding analysis in future research.
%However, our computational resources were limited, and we could not execute such an analysis. For the future research, it would be interesting to see results from such analysis.

\pagebreak[3] % Elastic pull from Reference
%\newpage

%\shrink \section{Acknowledgments}

%\begin{acks}
%To Robert, for the bagels and explaining CMYK and color spaces.
%\end{acks}

\bibliographystyle{ACM-Reference-Format}
\bibliography{sample-base}

% \appendix

\end{document}